\documentclass[10pt,letterpaper]{article}
\usepackage[top=0.85in,left=2.75in,footskip=0.75in]{geometry}

\usepackage{amsmath,amssymb}

\usepackage{changepage}

\usepackage[utf8x]{inputenc}

\usepackage{textcomp,marvosym}

\usepackage{cite}

\usepackage{nameref,hyperref}

\usepackage[right]{lineno}

\usepackage{microtype}
\DisableLigatures[f]{encoding = *, family = * }

\usepackage[table]{xcolor}

\usepackage{array}

\usepackage{enumerate}

\newcolumntype{+}{!{\vrule width 2pt}}

\newlength\savedwidth

\raggedright
\usepackage[aboveskip=1pt,labelfont=bf,labelsep=period,justification=raggedright,singlelinecheck=off]{caption}

\makeatletter
\renewcommand{\@biblabel}[1]{\quad#1.}
\makeatother

\date{}

\usepackage{lastpage,fancyhdr,graphicx}
\usepackage{epstopdf}
\fancyheadoffset[L]{2.25in}
\fancyfootoffset[L]{2.25in}
\newcommand{\withurl}[2]{{#1}\footnote{{\texttt{#2}}}}
\newcommand{\practicesection}[2]{\section{#1}\label{#2}}
\newcommand{\practice}[2]{\textbf{\emph{{#2}~({#1})}}}

\begin{document}
\vspace*{0.2in}

\begin{flushleft}
{\Large
\textbf\newline{Good Enough Practices in Scientific Computing}
}
\newline
\\
{Greg~Wilson}\textsuperscript{1,\ddag *},
{Jennifer~Bryan}\textsuperscript{2,\ddag},
{Karen~Cranston}\textsuperscript{3,\ddag},
{Justin~Kitzes}\textsuperscript{4,\ddag},
{Lex~Nederbragt}\textsuperscript{5,\ddag},
{Tracy~K.~Teal}\textsuperscript{6,\ddag}
\\
\textbf{1} Software Carpentry Foundation / gvwilson@software-carpentry.org
\\
\textbf{2} University of British Columbia / jenny@stat.ubc.ca
\\
\textbf{3} Duke University / karen.cranston@duke.edu
\\
\textbf{4} University of California, Berkeley / jkitzes@berkeley.edu
\\
\textbf{5} University of Oslo / lex.nederbragt@ibv.uio.no
\\
\textbf{6} Data Carpentry / tkteal@datacarpentry.org
\\
\bigskip
{\ddag} These authors contributed equally to this work.
\\
* E-mail: Corresponding gvwilson@software-carpentry.org

\end{flushleft}

\title{Good Enough Practices for Scientific Computing}

\section*{Abstract}

We present a set of computing tools and techniques that every
researcher can and should adopt.  These recommendations synthesize
inspiration from our own work, from the experiences of the thousands
of people who have taken part in Software Carpentry and Data Carpentry
workshops over the past six years, and from a variety of other guides.
Our recommendations are aimed specifically
at people who are new to research computing.

\section*{Author Summary}

Computers are now essential in all branches of science, but most
researchers are never taught the equivalent of basic lab skills for
research computing.  As a result, they take days or weeks to do things
that could be done in minutes or hours, are often unable to reproduce
their own work (much less the work of others), and have no idea how
reliable their computational results are.

This paper presents a set of good computing practices that every
researcher can adopt regardless of their current level of technical
skill.  These practices, which encompass data management, programming,
collaborating with colleagues, organizing projects, tracking work, and
writing manuscripts, are drawn from a wide variety of published
sources, from our daily lives, and from our work with volunteer
organizations that have delivered workshops to over 11,000 people
since 2010.

% \linenumbers

\section*{Introduction}\label{sec:introduction}

Two years ago a group of researchers involved in \withurl{Software
Carpentry}{http://software-carpentry.org/} and \withurl{Data
Carpentry}{http://datacarpentry.org/} wrote a paper called
``Best Practices for Scientific Computing'' \cite{wilson2014}.
It was well received, but many novices found its
litany of tools and techniques intimidating.  Also, by definition, the
``best'' are a small minority.
What practices are comfortably within reach for the ``rest''?

This paper therefore presents a set of ``good enough''
practices\footnote{Note that English lacks a good word for this:
  ``mediocre'', ``adequate'', and ``sufficient'' aren't exactly
  right.} for scientific computing, i.e., a minimum set of tools and
techniques that we believe every researcher can and should adopt. It
draws inspiration from many sources
\cite{gentzkow2014,noble2009,brown2015,wickham2014,kitzes2016,sandve2013,hart2015},
from our personal experience, and from the experiences of the
thousands of people who have taken part in Software Carpentry and
Data Carpentry workshops over the past six years.

Our intended audience is researchers who are working alone or with a
handful of collaborators on projects lasting a few days to a few
months, and who are ready to move beyond emailing themselves a spreadsheet
named \texttt{results-updated-3-revised.xlsx} at the end of the workday. A
practice is included in our list if large numbers of researchers use
it, and large numbers of people are \emph{still} using it months
after first trying it out. We include the second criterion because
there is no point recommending something that people won't actually
adopt.

Many of our recommendations are for the benefit of the collaborator
every researcher cares about most: their future self\footnote{As the
joke goes, yourself from three months ago doesn't answer
email{\ldots}}. Change is hard and if researchers don't see those benefits
quickly enough to justify the pain,
they will almost certainly switch back to their old way
of doing things.  This rules out many practices, such as code review,
that we feel are essential for larger-scale development
(Section~\ref{sec:omitted}).

We organize our recommendations into the following topics:

\begin{itemize}

\item Data Management:
  saving both raw and intermediate forms; documenting all steps; creating tidy data amenable to analysis.

\item Software:
  writing, organizing, and sharing scripts and programs used in an analysis.

\item Collaboration:
  making it easy for existing and new collaborators to understand and contribute to a project.

\item Project Organization:
  organizing the digital artifacts of a project to ease discovery and understanding.

\item Tracking Changes:
  recording how various components of your project change over time.

\item Manuscripts:
  writing manuscripts in a way that leaves an audit trail and minimizes manual merging of conflict.

\end{itemize}

\subsection*{Acknowledgments}

We are grateful to Arjun Raj (University of Pennsylvania), Steven
Haddock (Monterey Bay Aquarium Research Institute), Stephen Turner
(University of Virginia), Elizabeth Wickes (University of Illinois),
and Garrett Grolemund (RStudio) for their feedback on early versions
of this paper, to those who contributed during the outlining of the
manuscript, and to everyone involved in Data Carpentry and Software
Carpentry for everything they have taught us.

\practicesection{Data Management}{sec:data}

Data within a project may need to exist in various forms, ranging from
what first arrives to what is actually used for the primary analyses.
Our recommendations have two main themes. One is to work towards
ready-to-analyze data incrementally, documenting both the intermediate
data and the process. We also describe the key features of ``tidy
data'', which can be a powerful accelerator for analysis
\cite{wickham2014,hart2015}.

\begin{enumerate}

\item
  \practice{1a}{Save the raw data}.  Where possible,
  save data as originally generated (i.e. by an instrument or from a survey).
  It is
  tempting to overwrite raw data files with cleaned-up versions, but
  faithful retention is essential for re-running analyses
  from start to finish; for recovery from analytical
  mishaps; and for experimenting without fear. Consider changing file permissions to
  read-only or using spreadsheet protection features,
  so it is harder to damage raw
  data by accident or to hand edit it in a moment of weakness.

  Some data will be impractical to manage in this way. For example, you should avoid
  making local copies of large, stable repositories.
  In that case, record the exact procedure used to obtain the raw data, as well
  as any other pertinent info, such as an official version number or the date of download.

\item
  \practice{1b}{Create the data you wish to see in the world}. Create the dataset you
  \emph{wish} you had received.  The goal here is to improve machine and human
  readability, but \emph{not} to do vigorous data filtering or add external information.
  Machine readability allows automatic processing using computer programs,
  which is important when others want to reuse your data.
  Specific examples of non-destructive transformations that we recommend at the beginning
  of analysis:

  \emph{File formats}: Convert data from closed, proprietary formats to open,
  non-proprietary formats that ensure machine readability across time and computing
  setups \cite{ffIllinois}. Good options include CSV for tabular data, JSON, YAML, or XML for
  non-tabular data such as graphs\footnote{The node-and-arc kind.}, and HDF5 for certain
  kinds of structured data.

  \emph{Variable names}: Replace inscrutable variable names
  and artificial data codes with self-explaining alternatives, e.g.,
  rename variables called \texttt{name1} and \texttt{name2} to
  \texttt{personal\_name} and \texttt{family\_name}, recode the
  treatment variable from \texttt{1} vs.  \texttt{2} to
  \texttt{untreated} vs. \texttt{treated}, and replace artificial
  codes for missing data, such as ``-99'', with \texttt{NA}s,
  a code used in most programming languages to indicate that data is ``Not Available''
  \cite{white2013}.

  \emph{Filenames}: Store especially useful metadata as part of the filename itself, while
  keeping the filename regular enough for easy pattern matching. For example, a filename
  like \texttt{2016-05-alaska-b.csv} makes it easy for both people and programs to select
  by year or by location.

\item
  \practice{1c}{Create analysis-friendly data}: Analysis can be much easier if you are
  working with so-called ``tidy'' data \cite{wickham2014}. Two key principles are:

  \emph{Make each column a variable}: Don't cram two variables into one, e.g.,
  ``male\_treated'' should be split into separate variables for sex and treatment status.
  Store units in their own variable or in metadata, e.g., ``3.4'' instead of ``3.4kg''.

  \emph{Make each row an observation}: Data often comes in a wide format, because that
  facilitated data entry or human inspection. Imagine one row per field site and then
  columns for measurements made at each of several time points. Be prepared to gather such
  columns into a variable of measurements, plus a new variable for time point.
  Fig~\ref{fig:tidy} presents an example of such a transformation.

  \begin{figure} % \includegraphics[width = 5in]{tidy-data.png}
    \caption{\textbf{Example of gathering columns to create tidy data.}}
    \label{fig:tidy}
  \end{figure}

\item
  \practice{1d}{Record all the steps used to process data}: Data manipulation is as integral
  to your analysis as statistical modelling and inference. If you do not document this
  step thoroughly, it is impossible for you, or anyone else, to repeat the analysis.

  The best way to do this is to write scripts for \emph{every} stage of data processing.
  This might feel frustratingly slow, but you will get faster with practice. The
  immediate payoff will be the ease with which you can re-do data preparation when new
  data arrives. You can also re-use data preparation logic in the future for related
  projects.

  When scripting is not feasible, it's important to clearly document every manual action
  (what menu was used, what column was copied and pasted, what link was clicked, etc.).
  Often you can at least capture \emph{what} action was taken, if not the complete
  \emph{why}. For example, choosing a region of interest in an image is inherently
  interactive, but you can save the region chosen as a set of boundary coordinates.

\item
  \practice{1e}{Anticipate the need to use multiple tables}: Raw data, even if tidy, is not
  necessarily complete. For example, the primary data table might hold the heart rate for
  individual subjects at rest and after a physical challenge, identified via a subject ID.
  Demographic variables, such as subject age and sex, are stored in a second table and
  will need to be brought in via merging or lookup. This will go more smoothly if subject
  ID is represented in a common format in both tables, e.g., always as ``14025'' versus
  ``14,025'' in one table and ``014025'' in another. It is generally wise to give each
  record or unit a unique, persistent key and to use the same names and codes when
  variables in two datasets refer to the same thing.

\item
  \practice{1f}{Submit data to a reputable DOI-issuing repository so that
    others can access and cite it.}  Your data is as much a product of
  your research as the papers you write, and just as likely to be
  useful to others (if not more so).  Sites such as
  \withurl{Figshare}{https://figshare.com/},
  \withurl{Dryad}{http://datadryad.org/}, and
  \withurl{Zenodo}{https://zenodo.org/} allow others to find your work,
  use it, and cite it; we discuss licensing in
  Section~\ref{sec:collaboration} below. Follow your research community's
  standards for how to provide metadata. Note that there are two types
  of metadata: metadata about the dataset as a whole and metadata about
  the content within the dataset. If the audience is humans, write the
  metadata (the readme file) for humans. If the audience includes automatic
  metadata harvesters, fill out the formal metadata and write a good
  readme file for the humans \cite{wickes2015}.

\end{enumerate}

Taken in order, the recommendations above will produce intermediate
data files with increasing levels of cleanliness and
task-specificity. An alternative approach to data management would be
to fold all data management tasks into a monolithic procedure for data analysis,
so that intermediate data products are created ``on the fly'' and stored only in
memory, not saved as distinct files.

While the latter approach may be appropriate for projects in which very
little data cleaning or processing is needed, we recommend the explicit
creation and retention of intermediate products. Saving intermediate
files makes it easy to re-run \emph{parts} of a data analysis pipeline,
which in turn makes it less onerous to revisit and improve specific data
processing tasks. Breaking a lengthy workflow into pieces makes it
easier to understand, share, describe, and modify.

\practicesection{Software}{sec:software}

If you or your group are creating tens of thousands of lines of
software for use by hundreds of people you have never met, you are
doing software engineering. If you're writing a few dozen lines now
and again, and are probably going to be its only user, you may not be
doing engineering, but you can still make things easier on yourself by
adopting a few key engineering practices. What's more, adopting these
practices will make it easier for people to understand and (re)use
your code.

The core realization in these practices is that \emph{readable},
\emph{reusable}, and \emph{testable} are all side effects of writing
\emph{modular} code, i.e., of building programs out of short,
single-purpose functions with clearly-defined inputs and outputs.

\begin{enumerate}

\item
  \practice{2a}{Place a brief explanatory comment at the start of every
    program}, no matter how short it is. That comment should include
  at least one example of how the program is used: remember, a good
  example is worth a thousand words. Where possible, the comment
  should also indicate reasonable values for parameters.
  An example of such a comment is show below.

{\small
\begin{verbatim}
Synthesize image files for testing circularity estimation algorithm.

Usage: make_images.py -f fuzzing -n flaws -o output -s seed -v -w size

where:
-f fuzzing = fuzzing range of blobs (typically 0.0-0.2)
-n flaws   = p(success) for geometric distribution of # flaws/sample (e.g. 0.5-0.8)
-o output  = name of output file
-s seed    = random number generator seed (large integer)
-v         = verbose
-w size    = image width/height in pixels (typically 480-800)
\end{verbatim}
}

\item
  \practice{2b}{Decompose programs into functions} that are no more than
  one page (about 60 lines) long, do not
  use global variables (constants are OK), and take no more than half
  a dozen parameters.  The key motivation here is to fit the program
  into the most limited memory of all: ours. Human short-term memory
  is famously incapable of holding more than about seven items at
  once \cite{miller1956}. If we are to understand
  what our software is doing, we must break it into chunks that obey
  this limit, then create programs by combining these chunks.

\item
  \practice{2c}{Be ruthless about eliminating duplication}. Write and
  re-use functions instead of copying and pasting source code, and use
  data structures like lists rather than creating lots of variables
  called \texttt{score1}, \texttt{score2}, \texttt{score3}, etc.

  The easiest code to debug and maintain is code you didn't
  actually write, so \practice{2d}{always search for well-maintained
  software libraries that do what you need} before writing new code yourself,
  and \practice{2e}{test libraries before relying on them}.

\item
  \practice{2f}{Give functions and variables meaningful names}, both
  to document their purpose and to make the program easier to read. As
  a rule of thumb, the greater the scope of a variable, the more
  informative its name should be: while it's acceptable to call the
  counter variable in a loop \texttt{i} or \texttt{j}, the major data
  structures in a program should \emph{not} have one-letter
  names.  Remember to follow each language's conventions for names,
  such as \texttt{net\_charge} for Python and \texttt{NetCharge} for
  Java.

  \begin{quote}
    \noindent \textbf{Tab Completion}
    \\
    Almost all modern text editors provide \emph{tab completion}, so
    that typing the first part of a variable name and then pressing
    the tab key inserts the completed name of the variable.  Employing
    this means that meaningful longer variable names are no harder to type
    than terse abbreviations.

  \end{quote}

\item
  \practice{2g}{Make dependencies and requirements explicit.} This is
  usually done on a per-project rather than per-program basis, i.e.,
  by adding a file called something like \texttt{requirements.txt} to
  the root directory of the project, or by adding a ``Getting
  Started'' section to the \texttt{README} file.

\item
  \practice{2h}{Do not comment and uncomment sections of code to control
    a program's behavior}, since this is error prone and makes it
  difficult or impossible to automate analyses. Instead, put if/else
  statements in the program to control what it does.

\item
  \practice{2i}{Provide a simple example or test data set} that users
  (including yourself) can run to determine whether the program is
  working and whether it gives a known correct output for a
  simple known input. Such a ``build and smoke test'' is particularly
  helpful when supposedly-innocent changes are being made to the
  program, or when it has to run on several different machines, e.g.,
  the developer's laptop and the department's cluster.

\item
  \practice{2j}{Submit code to a reputable DOI-issuing repository} upon
  submission of paper, just as you do with data. Your software is as
  much a product of your research as your papers, and should be as
  easy for people to credit. DOIs for software are provided by
  \withurl{Figshare}{https://figshare.com/} and
  \withurl{Zenodo}{https://zenodo.org/}. Zenodo integrates directly
  with GitHub.

\end{enumerate}

\practicesection{Collaboration}{sec:collaboration}

You may start working on projects by yourself or with a small group of
collaborators you already know, but you should design it to make it easy
for new collaborators to join. These collaborators might be new
grad students or postdocs in the lab, or they might be \emph{you}
returning to a project that has been idle for some time. As summarized
in \cite{steinmacher2015}, you want to make it easy for people to set up a
local workspace so that they \emph{can} contribute, help them find tasks
so that they know \emph{what} to contribute, and make the contribution
process clear so that they know \emph{how} to contribute.  You also want
to make it easy for people to give you credit for your work.

\begin{enumerate}

\item
  \practice{3a}{Create an overview of your project.}  Have a short
  file in the project's home directory that explains the
  purpose of the project.  This file (generally called
  \texttt{README}, \texttt{README.txt}, or something similar) should
  contain the project's title, a brief description, up-to-date contact
  information, and
  an example or two of how to run various cleaning or analysis tasks.  It is often
  the first thing users of your project will look at, so make it
  explicit that you welcome contributors and point them to ways they
  can help.

  You should also create a \texttt{CONTRIBUTING} file that describes
  what people need to do in order to get the project going and
  contribute to it, i.e., dependencies that need to be installed,
  tests that can be run to ensure that software has been installed
  correctly, and guidelines or checklists that your project adheres
  to.

\item
  \practice{3b}{Create a shared public ``to-do'' list}.  This can be a
  plain text file called something like \texttt{notes.txt} or
  \texttt{todo.txt}, or you can use sites such as GitHub or Bitbucket
  to create a new \emph{issue} for each to-do item. (You can even add
  labels such as ``low hanging fruit'' to point newcomers at issues
  that are good starting points.)  Whatever you choose, describe the
  items clearly so that they make sense to newcomers.

\item
  \practice{3c}{Make the license explicit.}  Have a \texttt{LICENSE} file
  in the project's home directory that clearly states what license(s)
  apply to the project's software, data, and manuscripts. Lack of an
  explicit license does not mean there isn't one; rather, it implies
  the author is keeping all rights and others are not allowed to
  re-use or modify the material.

  We recommend Creative Commons licenses for data and text, either
  \withurl{CC-0}{https://creativecommons.org/about/cc0/} (the ``No
  Rights Reserved'' license) or
  \withurl{CC-BY}{https://creativecommons.org/licenses/by/4.0/} (the
  ``Attribution'' license, which permits sharing and reuse but requires people
  to give appropriate credit to the creators).  For software, we
  recommend a permissive open source license such as the MIT, BSD, or Apache
  license \cite{laurent2004}.

  \begin{quote}
    \noindent \textbf{What Not To Do}
    \\
    We recommend \emph{against} the ``no commercial use'' variations
    of the Creative Commons licenses because they may impede some
    forms of re-use.  For example, if a researcher in a developing
    country is being paid by her government to compile a public health
    report, she will be unable to include your data if the license says
    ``non-commercial''. We recommend permissive software licenses rather
    than the GNU General Public License (GPL) because it is easier to
    integrate permissively-licensed software into other projects,
    see chapter three in \cite{laurent2004}.
  \end{quote}

\item
  \practice{3d}{Make the project citable} by including a
  \texttt{CITATION} file in the project's home directory that
  describes how to cite this project as a whole, and where to find (and how to cite)
  any data sets, code, figures, and other artifacts
  that have their own DOIs.  The example below shows the
  \texttt{CITATION} file for the \withurl{Ecodata
    Retriever}{https://github.com/weecology/retriever}; for an example
  of a more detailed \texttt{CITATION} file, see the one for the
  \withurl{khmer}{https://github.com/dib-lab/khmer/blob/master/CITATION}
  project.

{\small
\begin{verbatim}
Please cite this work as:

Morris, B.D. and E.P. White. 2013. "The EcoData Retriever:
improving access to existing ecological data." PLOS ONE 8:e65848.
http://doi.org/doi:10.1371/journal.pone.0065848
\end{verbatim}
}

\end{enumerate}

\practicesection{Project Organization}{sec:project}

Organizing the files that make up a project in a logical and
consistent directory structure will help you and others keep track of
them.  Our recommendations for doing this are drawn primarily from
\cite{noble2009,gentzkow2014}.

\begin{enumerate}

\item
  \practice{4a}{Put each project in its own directory, which is named
    after the project.}  Like deciding when a chunk of code should be
  made a function, the ultimate goal of dividing research into
  distinct projects is to help you and others best understand your
  work. Some researchers create a separate project for each manuscript
  they are working on, while others group all research on a common
  theme, data set, or algorithm into a single project.

  As a rule of thumb, divide work into projects based on the overlap
  in data and code files. If two research efforts share no data or
  code, they will probably be easiest to manage independently. If they
  share more than half of their data and code, they are probably best
  managed together, while if you are building tools that are used in
  several projects, the common code should probably be in a project of
  its own.

\item
  \practice{4b}{Put text documents associated with the project in the
    \texttt{doc} directory.} This includes files for manuscripts,
  documentation for source code, and/or an electronic lab notebook
  recording your experiments.  Subdirectories may be created for these
  different classes of files in large projects.

\item
  \practice{4c}{Put raw data and metadata in a \texttt{data} directory,
    and files generated during cleanup and analysis in a
    \texttt{results} directory}, where ``generated files'' includes
  intermediate results, such as cleaned data sets or simulated data,
  as well as final results such as figures and tables.

  The \texttt{results} directory will \emph{usually}
  require additional subdirectories for all but the simplest
  projects. Intermediate files such as cleaned data, statistical
  tables, and final publication-ready figures or tables should be
  separated clearly by file naming conventions or placed into
  different subdirectories; those belonging to different papers or
  other publications should be grouped together.

\item
  \practice{4d}{Put project source code in the \texttt{src} directory.}
  \texttt{src} contains all of the code written for the project. This includes
  programs written in interpreted languages
  such as R or Python; those written compiled languages like
  Fortran, C++, or Java; as well as shell scripts, snippets of SQL used to pull
  information from databases; and other code needed to regenerate
  the results.

    This directory may contain two conceptually distinct types
    of files that should be distinguished either by clear file names or by
    additional subdirectories. The first type are files or
    groups of files that perform the core
    analysis of the research, such as data cleaning or statistical analyses.
    These files can be thought of as the
    ``scientific guts'' of the project.

    The second type of file in \texttt{src} is controller or driver
    scripts that combine the core analytical functions with particular
    parameters and data input/output commands in order to execute the
    entire project analysis from start to finish. A controller script for
    a simple project, for example, may read a raw data table, import and
    apply several cleanup and analysis functions from the other files in
    this directory, and create and save a numeric result. For a small
    project with one main output, a single controller script should be
    placed in the main \texttt{src} directory and distinguished clearly by
    a name such as ``runall''.

\item
  \practice{4e}{Put external scripts, or compiled programs
   in the \texttt{bin} directory}.  \texttt{bin} contains
  scripts that are brought in from elsewhere, and executable programs
  compiled from code in the \texttt{src} directory\footnote{The name
    \texttt{bin} is an old Unix convention, and comes from the term
    ``binary''.}. Projects that have neither will not require \texttt{bin}.

  \begin{quote}
    \noindent \textbf{Scripts vs.\ Programs}
    \\
    We use the term ``script'' to mean ``something that is executed
    directly as-is'', and ``program'' to mean ``something that is
    explicitly compiled before being used''.  The distinction is more
    one of degree than kind---libraries written in Python are actually
    compiled to bytecode as they are loaded, for example---so one
    other way to think of it is ``things that are edited directly''
    and ``things that are not''.
  \end{quote}

\item
  \practice{4f}{Name all files to reflect their content or function.} For
  example, use names such as \texttt{bird\_count\_table.csv},
  \texttt{manuscript.md}, or \texttt{sightings\_analysis.py}.  Do
  \emph{not} using sequential numbers (e.g., \texttt{result1.csv},
  \texttt{result2.csv}) or a location in a final manuscript (e.g.,
  \texttt{fig\_3\_a.png}), since those numbers will almost certainly
  change as the project evolves.

\end{enumerate}

The diagram below provides a concrete example of how a
simple project might be organized following these recommendations:

{\small
\begin{verbatim}
.
|-- CITATION
|-- README
|-- LICENSE
|-- requirements.txt
|-- data
|   -- birds_count_table.csv
|-- doc
|   -- notebook.md
|   -- manuscript.md
|   -- changelog.txt
|-- results
|   -- summarized_results.csv
|-- src
|   -- sightings_analysis.py
|   -- runall.py
\end{verbatim}
}

The root directory contains a \texttt{README} file that provides an
overview of the project as a whole, a \texttt{CITATION} file that
explains how to reference it, and a \texttt{LICENSE} file that states the
licensing. The \texttt{data} directory contains a
single CSV file with tabular data on bird counts (machine-readable
metadata could also be included here). The \texttt{src} directory
contains \texttt{sightings\_analysis.py}, a Python file containing
functions to summarize the tabular data, and a controller script
\texttt{runall.py} that loads the data table, applies functions
imported from \texttt{sightings\_analysis.py}, and saves a table of
summarized results in the \texttt{results} directory.

This project doesn't have a \texttt{bin} directory, since it does not
rely on any compiled software. The \texttt{doc} directory contains two
text files written in Markdown, one containing a running lab notebook
describing various ideas for the project and how these were
implemented and the other containing a running draft of a manuscript
describing the project findings.

\practicesection{Keeping Track of Changes}{sec:versioning}

Keeping track of changes that you or your collaborators make to data
and software is a critical part of research. Being able to reference or
retrieve a specific version of the entire project aids in reproducibility
for you leading up to publication, when responding to reviewer comments,
and when providing supporting information for reviewers, editors,
and readers.

We believe that the best tools for tracking changes are the version
control systems that are used in software development, such as Git,
Mercurial, and Subversion. They keep track of what was changed in a
file when and by whom, and synchronize changes to a central server so
that many users can manage changes to the same set of files.

Although all of the authors use version control daily for all of their
projects, we recognize that many newcomers to computational science
find version control to be one of the more difficult practices to
adopt.  We therefore recommend that projects adopt \emph{either} a
systematic manual approach for managing changes \emph{or} version
control in its full glory.

Whatever system you chose, we recommend that you use it in the following way:

\begin{enumerate}

\item
  \practice{5a}{Back up (almost) everything created by a human being as
    soon as it is created.} This includes scripts and programs of all
  kinds, software packages that your project depends on, and
  documentation. A few exceptions to this rule are discussed below.

\item
  \practice{5b}{Keep changes small.}  Each change should not be so large
  as to make the change tracking irrelevant. For example, a single
  change such as ``Revise script file'' that adds or changes several
  hundred lines is likely too large, as it will not allow changes to
  different components of an analysis to be investigated
  separately. Similarly, changes should not be broken up into pieces
  that are too small. As a rule of thumb, a good size for a single change is
  a group of edits that you could imagine wanting to undo in one step
  at some point in the future.

\item
  \practice{5c}{Share changes frequently.} Everyone working on the
  project should share and incorporate changes from others on a
  regular basis. Do not allow individual investigator's versions of
  the project repository to drift apart, as the effort required to
  merge differences goes up faster than the size of the
  difference. This is particularly important for the manual versioning
  procedure describe below, which does not provide any
  assistance for merging simultaneous, possibly conflicting, changes.

\item
  \practice{5d}{Create, maintain, and use a checklist for saving and
    sharing changes to the project.} The list should include writing
  log messages that clearly explain any changes, the size and content
  of individual changes, style guidelines for code, updating to-do
  lists, and bans on committing half-done work or broken code.  See
  \cite{gawande2011} for more on the proven value of checklists.

\item
  \practice{5e}{Store each project in a folder that is mirrored off the
    researcher's working machine} using a system such as Dropbox
    or a remote version control repository such as GitHub.
    Synchronize that folder at least daily. It may take a few minutes,
  but that time is repaid the
  moment a laptop is stolen or its hard drive fails.

\end{enumerate}

\subsection*{Manual Versioning}

Our first suggested approach, in which everything is done by hand, has
two additional parts:

\begin{enumerate}

\item
  \practice{5f}{Add a file called \texttt{CHANGELOG.txt} to the project's
    \texttt{docs} subfolder}, and make dated notes about changes to
  the project in this file in reverse chronological order (i.e., most
  recent first). This file is the equivalent of a lab notebook, and
  should contain entries like those shown below.

{\small
\begin{verbatim}
## 2016-04-08

* Switched to cubic interpolation as default.
* Moved question about family's TB history to end of questionnaire.

## 2016-04-06

* Added option for cubic interpolation.
* Removed question about staph exposure (can be inferred from blood test results).
\end{verbatim}
}

\item
  \practice{5g}{Copy the entire project whenever a significant change has
    been made} (i.e., one that materially affects the results), and store that
    copy in a sub-folder whose name reflects
  the date in the area that's being synchronized. This approach
  results in projects being organized as shown below:

{\small
\begin{verbatim}
.
|-- project_name
|   -- current
|       -- ...project content as described earlier...
|   -- 2016-03-01
|       -- ...content of 'current' on Mar 1, 2016
|   -- 2016-02-19
|       -- ...content of 'current' on Feb 19, 2016
\end{verbatim}
}

  Here, the \texttt{project\_name} folder is
  mapped to external storage (such as Dropbox), \texttt{current} is
  where development is done, and other folders within
  \texttt{project\_name} are old versions.

  \begin{quote}
    \noindent \textbf{Data is Cheap, Time is Expensive}
    \\
    Copying everything like this may seem wasteful, since many files
    won't have changed, but consider: a terabyte hard drive costs
    about \$50 retail, which means that 50 GByte costs less than a
    latte. Provided large data files are kept out of the backed-up
    area (discussed below), this approach costs less than the time it
    would take to select files by hand for copying.
  \end{quote}

\end{enumerate}

This manual procedure satisfies the requirements outlined above
without needing any new tools. If multiple researchers are working on
the same project, though, they will need to coordinate so that only a
single person is working on specific files at any time. In particular,
they may wish to create one change log file per contributor, and to
merge those files whenever a backup copy is made.

\subsection*{Version Control Systems}

What the manual process described above requires most is
self-discipline. The version control tools that underpin our second
approach---the one all authors use for their projects---don't just
accelerate the manual process: they also automate some steps while
enforcing others, and thereby require less self-discipline for more
reliable results.

\begin{enumerate}

\item
  \practice{5h}{Use a version control system}, to manage changes
  to a project.

\end{enumerate}

Box~2 briefly explains how version control systems work.
It's hard to know what version control tool is most widely used in
research today, but the one that's most talked about is undoubtedly
\withurl{Git}{https://git-scm.com/}. This is largely because of
\withurl{GitHub}{http://github.com}, a popular hosting site that
combines the technical infrastructure for collaboration via Git with
a modern web interface. GitHub is free for public and open source
projects and for users in academia and nonprofits.
\withurl{GitLab}{https://about.gitlab.com} is a well-regarded
alternative that some prefer, because the GitLab platform itself is
free and open source.

For those who find Git's command-line syntax inconsistent
and confusing, \withurl{Mercurial}{https://www.mercurial-scm.org/} is
a good choice; \withurl{Bitbucket}{https://bitbucket.org/} provides
free hosting for both Git and Mercurial repositories, but does not
have nearly as many scientific users.

\subsection*{What Not to Put Under Version Control}

The benefits of version control systems don't apply equally to all
file types.  In particular, version control can be more or less
rewarding depending on file size and format.  First, file comparison
in version control systems is optimized for plain text files, such as
source code. The ability to see so-called ``diffs'' is one of the
great joys of version control. Unfortunately, Microsoft Office files
(like the \texttt{.docx} files used by Word) or other binary files,
e.g., PDFs, can be stored in a version control system, but it is not
possible to pinpoint specific changes from one version to the next.
Tabular data (such as CSV files) can be put in version control, but
changing the order of the rows or columns will create a big change for
the version control system, even if the data itself has not changed.

Second, raw data should not change, and therefore should not require
version tracking.  Keeping intermediate data files and other results
under version control is also not necessary if you can re-generate
them from raw data and software. However, if data and results are
small, we still recommend versioning them for ease of access by
collaborators and for comparison across versions.

Third, today's version control systems are not designed to handle
megabyte-sized files, never mind gigabytes, so large data or results
files should not be included.  (As a benchmark for ``large'', the
limit for an individual file on GitHub is 100MB.)  Some emerging
hybrid systems such as \withurl{Git LFS}{https://git-lfs.github.com/}
put textual notes under version control, while storing the large data
itself in a remote server, but these are not yet mature enough for us
to recommend.

\begin{quote}
  \noindent \textbf{Inadvertent Sharing}
  \\
  Researchers dealing with data subject to legal restrictions that
  prohibit sharing (such as medical data) should be careful not to put
  data in public version control systems. Some institutions may
  provide access to private version control systems, so it is worth
  checking with your IT department.
  \\
  Additionally, be sure not to unintentionally place security credentials,
  such as passwords and private keys, in a version control system where it
  may be accessed by others.
\end{quote}

\practicesection{Manuscripts}{sec:manuscripts}

An old joke says that doing the research is the first 90\% of any
project; writing up is the other 90\%. While writing is rarely
addressed in discussions of scientific computing, computing has
changed scientific writing just as much as it has changed research.

A common practice in academic writing is for the lead author to send
successive versions of a manuscript to coauthors to collect feedback,
which is returned as changes to the document, comments on the
document, plain text in email, or a mix of all three. This results in
a lot of files to keep track of, and a lot of tedious manual labor to
merge comments to create the next master version.

Instead of an email-based workflow, we recommend mirroring good
practices for managing software and data to make writing scalable,
collaborative, and reproducible.  As with our recommendations for
version control in general, we suggest that groups choose one of two
different approaches for managing manuscripts.  The goals of both are
to:

\begin{itemize}

\item
  Ensure that text is accessible to yourself and others now and in the
  future by making a single master document that is available to all
  coauthors at all times.

\item
  Reduce the chances of work being lost or people overwriting each
  other's work.

\item
  Make it easy to track and combine contributions from multiple
  collaborators.

\item
  Avoid duplication and manual entry of information, particularly in
  constructing bibliographies, tables of contents, and lists.

\item
  Make it easy to regenerate the final published form (e.g., a PDF)
  and to tell if it is up to date.

\item
  Make it easy to share that final version with collaborators and to
  submit it to a journal.

\end{itemize}

\begin{quote}
  \noindent \textbf{The First Rule Is{\ldots}}
  \\
  The workflow you choose is less important than having all authors
  agree on the workflow \emph{before} writing starts. Make sure to
  also agree on a single method to provide feedback, be it an email
  thread or mailing list, an issue tracker (like the ones provided by
  GitHub and Bitbucket), or some sort of shared online to-do list.
\end{quote}

\subsection*{Single Master Online}

Our first alternative will already be familiar to many researchers:

\begin{enumerate}

\item
  \practice{6a}{Write manuscripts using online tools with rich
    formatting, change tracking, and reference management}, such as
  Google Docs. With the document online, everyone's
  changes are in one place, and hence don't need to be merged
  manually.

% \item
%   \practice{6b}{Include a \texttt{PUBLICATIONS} file in the project's
%     \texttt{doc} directory} with metadata about each online manuscript
%   (e.g., their URLs). This is analogous to the \texttt{data}
%   directory, which might contain links to the location of the data
%   file(s) rather than the actual files.

\end{enumerate}

We realize that in many cases, even this solution is asking too much
from collaborators who see no reason to move forward from desktop GUI
tools. To satisfy them, the manuscript can be converted to a desktop
editor file format (e.g., Microsoft Word's \texttt{.docx} or
LibreOffice's \texttt{.odt}) after major changes, then downloaded and
saved in the \texttt{doc} folder. Unfortunately, this means merging
some changes and suggestions manually, as existing tools cannot always
do this automatically when switching from a desktop file format to
text and back (although \withurl{Pandoc}{http://pandoc.org/} can go a
long way).

\subsection*{Text-based Documents Under Version Control}

The second approach treats papers exactly like software, and has been
used by researchers in mathematics, astronomy, physics, and related
disciplines for decades:

\begin{enumerate}

\item
  \practice{6b}{Write the manuscript in a plain text format that permits
    version control} such as
  \withurl{LaTeX}{http://www.latex-project.org/} or
  \withurl{Markdown}{http://daringfireball.net/projects/markdown/},
  and then convert them to other formats such as PDF as needed using
  scriptable tools like \withurl{Pandoc}{http://pandoc.org/}.

% \item
%   \practice{6d}{Include tools needed to compile manuscripts in the
%     project folder} and keep them under version control just like
%   tools used to do simulation or analysis.

\end{enumerate}

Using a version control system provides good support for finding
and merging differences resulting from concurrent changes. It
also provides a convenient platform for making comments and
performing review.

This approach re-uses the version control tools and skills used to
manage data and software, and is a good starting point for
fully-reproducible research. However, it requires all contributors to
understand a much larger set of tools, including markdown or LaTeX,
make, BiBTeX, and Git/GitHub.

\subsection*{Why Two Recommendations for Manuscripts?}

The first draft of this paper recommended always using plain text in
version control to manage manuscripts, but several reviewers pushed
back forcefully. For example, Stephen Turner wrote:

\begin{quote}
{\ldots}try to explain the notion of compiling a document to an
overworked physician you collaborate with. Oh, but before that, you
have to explain the difference between plain text and word
processing. And text editors. And markdown/LaTeX compilers. And
BiBTeX. And Git. And GitHub. Etc. Meanwhile he/she is getting paged
from the OR{\ldots}

{\ldots}as much as we want to convince ourselves otherwise, when you
have to collaborate with those outside the scientific computing
bubble, the barrier to collaborating on papers in this framework is
simply too high to overcome. Good intentions aside, it always comes
down to ``just give me a Word document with tracked changes,'' or
similar.
\end{quote}

Similarly, Arjun Raj said in \withurl{a blog
  post}{http://rajlaboratory.blogspot.ca/2016/03/from-over-reproducibility-to.html}:

\begin{quote}
Google Docs excels at easy sharing, collaboration, simultaneous
editing, commenting and reply-to-commenting. Sure, one can approximate
these using text-based systems and version control. The question is
why anyone would like to{\ldots}

The goal of reproducible research is to make sure one
can{\dots}reproduce{\ldots}computational analyses. The goal of version
control is to track changes to source code. These are fundamentally
distinct goals, and while there is some overlap, version control is
merely a tool to help achieve that, and comes with so much overhead
and baggage that it is often not worth the effort.
\end{quote}

In keeping with our goal of recommending ``good enough'' practices, we
have therefore included online, collaborative editing in something like Google
Docs. We still recommend \emph{against} traditional desktop tools like
LibreOffice and Microsoft Word because they make collaboration more
difficult than necessary.

\subsection*{Supplementary Materials}

Supplementary materials often contain much of the work that went into
the project, such as tables and figures or more elaborate descriptions
of the algorithms, software, methods, and analyses. In order to make
these materials as accessible to others as possible, do not rely
solely on the PDF format, since extracting data from PDFs is
notoriously hard.  Instead, we recommend separating the results that
you may expect others to reuse (e.g., data in tables, data behind
figures) into separate, text-format files in formats such as
CSV, JSON, YAML, XML, or HDF5\footnote{We recommend against more
innovative formats in deference to an old saying: ``What's oldest lasts longest.''}.
The same holds for any commands or code you want to include as
supplementary material: use the format that most easily enables reuse
(source code files, Unix shell scripts etc).

\practicesection{What We Left Out}{sec:omitted}

We have deliberately left many good tools and practices off our list,
including some that we use daily, because they only make sense on top
of the core practices described above, or because it takes a larger
investment before they start to pay off.

\begin{description}

\item[\textbf{Branches}] A \emph{branch} is a ``parallel universe''
  within a version control repository. Developers create branches so
  that they can make multiple changes to a project independently. They
  are central to the way that experienced developers use systems like
  Git, but they add an extra layer of complexity to version control
  for newcomers.  Programmers got along fine in the days of CVS and
  Subversion without relying heavily on branching, and branching can
  be adopted without significant disruption after people have mastered
  a basic edit-commit workflow.

\item[\textbf{Build Tools}] Tools like
  \withurl{Make}{https://www.gnu.org/software/make/} were originally
  developed to recompile pieces of software that had fallen out of
  date. They are now used to regenerate data and entire papers: when
  one or more raw input files change, Make can automatically re-run
  those parts of the analysis that are affected, regenerate tables and
  plots, and then regenerate the human-readable PDF that depends on
  them.  However, newcomers can achieve the same behavior by writing
  shell scripts that re-run everything; these may do unnecessary work,
  but given the speed of today's machines, that is unimportant for
  small projects.

\item[\textbf{Unit Tests}] A \emph{unit test} is a small test of one
  particular feature of a piece of software. Projects rely on unit
  tests to prevent \emph{regression}, i.e., to ensure that a change to
  one part of the software doesn't break other parts. While unit tests
  are essential to the health of large libraries and programs, we have
  found that they usually aren't compelling for solo exploratory
  work. (Note, for example, the lack of a \texttt{test} directory in
  Noble's rules \cite{noble2009}.)  Rather than advocating something
  which people are unlikely to adopt, we have left unit testing off
  this list.

\item[\textbf{Continuous Integration}] Tools like
  \withurl{Travis-CI}{https://travis-ci.org/} automatically run a set
  of user-defined commands whenever changes are made to a version
  control repository. These commands typically execute tests to make
  sure that software hasn't regressed, i.e., that things which used to
  work still do. These tests can be run either before changes are
  saved (in which case the changes can be rejected if something fails)
  or after (in which case the project's contributors can be notified
  of the breakage). CI systems are invaluable in large projects with
  many contributors, but pay fewer dividends in smaller projects where
  code is being written to do specific analyses.

\item[\textbf{Profiling and Performance Tuning}] \emph{Profiling} is
  the act of measuring where a program spends its time, and is an
  essential first step in \emph{tuning} the program (i.e., making it
  run faster). Both are worth doing, but only when the program's
  performance is actually a bottleneck: in our experience, most users
  spend more time getting the program right in the first place.

\item[\textbf{Coverage}] Every modern programming language comes with
  tools to report the \emph{coverage} of a set of test cases, i.e.,
  the set of lines that are and aren't actually executed when those
  tests are run. But as with unit testing, this
  only starts to pay off once the project grows larger, and is
  therefore not recommended here.

\item[\textbf{The Semantic Web}] Ontologies and other formal
  definitions of data are useful, but in our experience, even
  simplified things like \withurl{Dublin Core}{http://dublincore.org/}
  are rarely encountered in the wild.

\item[\textbf{Documentation}] Good documentation is a key factor in
  software adoption, but in practice, people won't write comprehensive
  documentation until they have collaborators who will use it. They
  will, however, quickly see the point of a brief explanatory comment
  at the start of each script, so we have recommended that as a first
  step.

\item[\textbf{A Bibliography Manager}] Researchers should use a
  reference manager of some sort, such as
  \withurl{Zotero}{http://zotero.org/}, and should also obtain and use
  an \withurl{ORCID}{http://orcid.org/} to identify themselves in
  their publications, but discussion of those is outside the scope of
  this paper.

\item[\textbf{Code Reviews and Pair Programming}] These practices are
  valuable in projects with multiple people making large software
  contributions, which is not typical for the intended audience for this paper \cite{petre2014}.

\end{description}

One important observation about this list is that many experienced
programmers actually do some or all of these things even for small
projects. It makes sense for them to do so because (a) they've already
paid the learning cost of the tool, so the time required to implement
for the ``next'' project is small, and (b) they understand that their
project will need some or all of these things as it scales, so they
might as well put it in place now.

The problem comes when those experienced developers give advice to
people who \emph{haven't} already mastered the tools, and
\emph{don't} realize (yet) that they will save time if and when their
project grows.  In that situation, advocating unit testing with
coverage checking and continuous integration is more likely to scare
newcomers off than to aid them.

\section*{Conclusion}

We have outlined a series of practices for scientific computing based
on our collective experience, and the experience of the thousands of
researchers we have met through Software Carpentry, Data Carpentry,
and similar organizations.  These practices are pragmatic, accessible
to people who consider themselves computing novices, and can be applied
by both individuals and groups.
Most importantly, these practices make researchers more productive
individually by enabling them to get more done in less time and with
less pain.  They also accelerate research as a whole by making
computational work (which increasingly means \emph{all} work) more
reproducible.

But progress will not happen by itself.  Universities and funding
agencies need to support training for researchers in the use of these
tools.  Such investment will improve confidence in the results of
computational work and allow us to make more rapid progress on
important research questions.

% \bibliography{good-enough-practices-for-scientific-computing}

\pagebreak

\section*{Box 1: Summary of Practices}

{\footnotesize
\begin{enumerate}
\item Data Management
  \begin{enumerate}[a)]
  \item Save the raw data.
  \item Create the data you wish to see in the world.
  \item Create analysis-friendly data.
  \item Record all the steps used to process data.
  \item Anticipate the need to use multiple tables.
  \item Submit data to a reputable DOI-issuing repository so that others can access and cite it.
  \end{enumerate}
\item Software
  \begin{enumerate}[a)]
  \item Place a brief explanatory comment at the start of every program.
  \item Decompose programs into functions.
  \item Be ruthless about eliminating duplication.
  \item Always search for well-maintained software libraries that do what you need.
  \item Test libraries before relying on them.
  \item Give functions and variables meaningful names.
  \item Make dependencies and requirements explicit.
  \item Do not comment and uncomment sections of code to control a program's behavior.
  \item Provide a simple example or test data set.
  \item Submit code to a reputable DOI-issuing repository.	
  \end{enumerate}
\item Collaboration
  \begin{enumerate}[a)]
  \item Create an overview of your project.
  \item Create a shared public ``to-do'' list.
  \item Make the license explicit.
  \item Make the project citable.
  \end{enumerate}
\item Project Organization
  \begin{enumerate}[a)]
  \item Put each project in its own directory, which is named after the project.
  \item Put text documents associated with the project in the \texttt{doc} directory.
  \item Put raw data and metadata in a \texttt{data} directory, and files generated during cleanup and analysis in a \texttt{results} directory.
  \item Put project source code in the \texttt{src} directory.
  \item Put external scripts, or compiled programs in the \texttt{bin} directory.
  \item Name all files to reflect their content or function.
  \end{enumerate}
\item Keeping Track of Changes
  \begin{enumerate}[a)]
  \item Back up (almost) everything created by a human being as soon as it is created.
  \item Keep changes small.
  \item Share changes frequently.
  \item Create, maintain, and use a checklist for saving and sharing changes to the project.
  \item Store each project in a folder that is mirrored off the researcher's working machine.
   \item Use a file called \texttt{CHANGELOG.txt} to record changes, and
   \item Copy the entire project whenever a significant change has been made, OR
   \item Use a version control system to manage changes 
  \end{enumerate}
\item Manuscripts
  \begin{enumerate}[a)]
  \item Write manuscripts using online tools with rich formatting, change tracking, and reference management, OR
  \item Write the manuscript in a plain text format that permits version control
  % \item Include a \texttt{PUBLICATIONS} file in the project's \texttt{doc} directory.
  % \item Include any tools needed to compile manuscripts in the project folder.
  \end{enumerate}
\end{enumerate}
}

\pagebreak

\section*{Box 2: How Version Control Systems Work}

A version control system stores snapshots of a project's files in a
repository. Users modify their working copy of the project, and then
save changes to the repository when they wish to make
a permanent record and/or share their work with colleagues. The
version control system automatically records when the change was
made and by whom along with the changes themselves.

Crucially, if several people have edited files simultaneously, the
version control system will detect the collision and require them to
resolve any conflicts before recording the changes. Modern version
control systems also allow repositories to be synchronized with each
other, so that no one repository becomes a single point of failure.
Tool-based version control has several benefits over manual version
control:

\begin{itemize}

\item
  Instead of requiring users to make backup copies of the whole project,
  version control safely stores just enough information to allow old
  versions of files to be re-created on demand.

\item
  Instead of relying on users to choose sensible names for backup
  copies, the version control system timestamps all saved changes
  automatically.

\item
  Instead of requiring users to be disciplined about completing the
  changelog, version control systems prompt them every time a change
  is saved. They also keep a 100\% accurate record of what was
  \emph{actually} changed, as opposed to what the user
  \emph{thought} they changed, which can be invaluable when problems
  crop up later.

\item
  Instead of simply copying files to remote storage, version control
  checks to see whether doing that would overwrite anyone else's
  work. If so, they facilitate identifying conflict and merging changes.

\end{itemize}

\pagebreak

\end{document}